\documentclass{article}

\usepackage{arxiv}

\usepackage[T1]{fontenc}
\usepackage{hyperref}       
\usepackage{url}            
\usepackage{booktabs}       
\usepackage{amsfonts}       
\usepackage{nicefrac}       
\usepackage{microtype}      
\usepackage{lipsum}
\usepackage{todonotes}
\usepackage{booktabs}
\usepackage{multirow}
\usepackage{amsmath}
\usepackage{placeins}
\usepackage{cleveref}

\newcommand\blfootnote[1]{%
  \begingroup
  \renewcommand\thefootnote{}\footnote{#1}%
  \addtocounter{footnote}{-1}%
  \endgroup
}

\title{Mind The Gap: Real-time Decentralized Distance Estimation using Ultrasound and Bluetooth across Multiple Smartphones}

\author{
Devansh~R.~Agrawal \thanks{Department of Aeronautics, Imperial College London. \texttt{devansh@hackpartners.com}}\\
Hack Partners Limited, London
\And
Peter~Lyon \thanks{Yak Consultancy Ltd, Bulgaria. \texttt{peter@yakconsultancy.com}}\\
Hack Partners Limited, London\\
via Yak Consultancy
\And
Martin~Frobisher\\
Network Rail, London
\And
Andy~Doherty\\
Network Rail, London
\And
Ben~Allen\\
Network Rail, London
\And
Freddie~Rawlins \thanks{Department of Computer Science, Oxford University. \texttt{freddie@hackpartners.com}}\\
Hack Partners Limited, London
}

\begin{document}
\blfootnote{Permission to make digital or hard copies of all or part of this work for personal use is granted without fee provided that copies are not made or distributed for profit or commercial advantage and that copies bear this notice on the first page. To copy otherwise, or republish, to post on servers or to redistribute to lists, requires prior specific permission and/or fee. Request permissions from \texttt{mindthegap@hackpartners.com}. No transfer, grant or license of rights under any patent or copyright or to any intellectual property, proprietary information and/or trade secret is made or is to be implied by this notice.
\copyright 2020 Hack Partners Limited. All rights reserved.}
\blfootnote{Please send all correspondence and questions to River Baig, \texttt{river@hackpartners.com}}
\maketitle

\begin{abstract}

Robust, low-cost solutions are needed to maintain social distancing guidelines during the COVID-19 pandemic. We establish a method to measure the distance between multiple phones across a large number of closely spaced smartphones with a median absolute error of 8.5~cm. The application works in real-time, using Time of Flight of near-ultrasound signals, providing alerts with sufficient responsiveness to be useful for distancing while devices are in users pockets and they are moving at walking speed. The approach is decentralized, requires no additional hardware, and can operate in the background without an internet connection. We have no device specific requirements nor need any manual calibration or device synchronization. It has been tested with over 20 different phones models, from both the Android and iOS systems in the past 5 years.
To the best of our knowledge, this is the first successful such implementation, and has 25000 users at time of publishing.
\end{abstract}

\keywords{Indoor positioning \and Smartphone-based Positioning \and Real-time Distance Measurements \and COVID-19 \and Social Distancing}

\section{Introduction}

Due to the global spread of the COVID-19 pandemic, many governments worldwide implemented social-distancing guidelines to reduce the rate of transmission amongst the community \cite{WHO2020}. For instance, the UK government has indicated that individuals must remain 2~meters apart, or 1~meter if appropriate measures are in place \cite{UKGuidance}. 

Mind The Gap is a smartphone application developed in response to these requirements. It uses the device's ultrasound\footnote{Technically, we are using near-ultrasound, but for brevity we refer to this simply as ultrasound.} and Bluetooth Low Energy capabilities to determine the distance to the closest neighboring smartphone that is running the app, and alert the user if they are closer than the user-specified distance.

In this paper, we present the fully decentralised, distributed algorithm and audio signal processing method that allows the application to operate in real-time and unstructured environments like offices and outdoors. Our system provides relative indoor distance measurements, with a median absolute error of 8.5~cm, despite obstructions and  multi-path effects. It requires no additional beacons or hardware, and does not require any calibration or  synchronization.  At the time of publication, the application has over 2000 daily active users, with positive feedback on the utility and accuracy of the system.

The paper is structured as follows. \Cref{sec:background} provides some background to the problem of real-time distance measurements and in \cref{sec:contributions} we list the key contributions of this paper. \Cref{sec:algorithm} describes the algorithm, its scalability and the signal processing approach to enable high frequency updates across large groups of individuals. Finally in \cref{sec:performance} we describe the performance of the application, including testing results from a third-party testing agency.

\subsection{Background}
\label{sec:background}

A variety of methods have been developed for indoor-positioning over a range of accuracy requirements \cite{MendozaSilva2019}. Global positioning, for instance GPS typically provides accuracy to 4.9~meters, and most indoor positioning systems provide accuracy in the order of centimeters to a meter. Mendoza-Silva categorized the principles into (1) Time of Arrival (TOA) of a signal from an emitter, (2) Time Difference of Arrival (TDOA) from multiple synchronised emitters, (3) Angle of Arrival relative to a set of known locations and (4) Received Signal Strength (RSS), which maps the RSS within the environment to determine the position in the future. 

For our application, we require infrastructure-free solutions that can operate on smartphones with no additional hardware, or prior environment mapping. It must operate without Line of Sight (LOS), for example through pockets or backpacks, and around the human body. Finally, since social distancing requires separations of 1~-~2 meters, our measurement accuracy must be one order better, approximately 10~cm. 

Of the 10 methods identified by Mendoza-Silva, including Light, Computer Vision, Wifi, and Bluetooth), ultrasound based methods meet our requirements \cite{MendozaSilva2019}. Wifi and Bluetooth would be based on either RSS (which correlates poorly with distance) or fingerprinting (which would require offline maps to be generated for all environments).

A number of ultrasound based approaches have been discussed before, including ``Active Bat'' \cite{Ward1997} ``Cricket" \cite{Cricket}, ``Dolphin'' \cite{Fukuju}, and ``BeepBeep'' \cite{Peng2007}. These systems have been able to achieve remarkable accuracy, upto 1~cm accuracy. However, these are achieved  only for one or a few devices of unknown position and with careful placement and calibration of multiple fixed position beacons. BeepBeep is the only implementation above which could measure the distance between a pair of smartphones without additional hardware, and achieved accuracy of 1-2~cm with 2~cm of standard deviation (after averaging a number of measurements). They used 50~millisecond long linear chirps, and communicated over WiFi \cite{Peng2007}. 

More recent implementations from the Microsoft Indoor Localization Competition demonstrated accuracy of of 10~cm, while employing beacons, and using encoded communication protocols including Hamming codes and Binary Phase Shift Keying \cite{Urena2018}. These authors also reported challenges with multi-path effects, with accuracy reducing to as much as 90~cm.

Our work builds on these ultrasound-based communication systems. This paper's key contributions are listed next.

\subsection{Contributions}
\label{sec:contributions}

\begin{itemize}
    \item Fully distributed and decentralised ranging architecture. This architecture allows for good scalability with number of devices and device density.
    \item A simple reflection and obstruction resistant very-short audio pulse design and detection algorithm. This system also works through clothing. 
    \item A fully-software solution that requires no specialised/additional hardware beyond a standard smart-phone. No calibration or mapping is necessary.
    \item The entire solution is optimized to limit battery consumption.
\end{itemize}

These claims are discussed and supported in the following sections.

\section{Algorithm Description}
\label{sec:algorithm}

Our solution must tackle a few key challenges: (1) it must work across large network of spread-out users, (2) it must allow for users to dynamically enter and leave a region, (3) it must be operational entirely in the background, even through pockets and (4) must have a high enough update frequency to account for people walking at normal speeds. Furthermore, we required the solution to work without a centralized coordinator, and no additional beacons at known locations. 

\subsection{Scaling Considerations}

To illustrate the scaling challenge, we will discuss an office-floor like scenario. First we establish terminology.

Two users are defined as being \emph{breaching}, if they are closer than a predefined tolerance $d=2$~m, \emph{close} if they are within Bluetooth and audio range of each of other and \emph{connected} if there is a path of `close' users between them (\cref{fig:ranging}). In practice, the physical range for two devices to be close is less than 3.5~meters. For an individual $i$, we define the set of breaching users as $B_i$, close users $C_i$, and connected users $X_i$, and $| \cdot | $ as the size of a set.  Note, $B_i \subseteq C_i \subseteq X_i$. Let the set $X = \bigcup_{i}X_i$ be the set of all connected users at some time. We also do not consider devices with a wall in between as breaching, as it is irrelevant to the COVID-19 scenario. 

To allow the system to operate efficiently for large $|X|$, we must ensure the maximum time to update the distance to the nearest neighbor, $T$ scales well with $|X|$. Letting $t$ be time required to measure the distance between a pair of users, simply computing pairwise distances across $X$ in series will require $T = t |X| (|X|-1)/2 $ seconds. Even for $t=0.5$~seconds, with $|X| = 30$, the time to update is $T=3.6$~minutes!

In a regular office setting, and particularly during periods of social distancing, while all users on a floor may be connected to each other, only~$|C_i|\approx5$~users may be close to individual~$i$. Therefore, if we can run distance updates on each set of close users in parallel, we scale with $T\sim\mathcal{O}(|C|^2t)$ instead, and $T\approx 6$~seconds.

We can do even better, scaling as $T\sim|C|t/2$, as is discussed in the next section. This allows the distances across the entire group of phones to be update once every 1-1.5~seconds. 

\subsection{Decentralized Parallel Ranging}

\begin{figure}[tb]
    \centering
    \includegraphics[width=\linewidth]{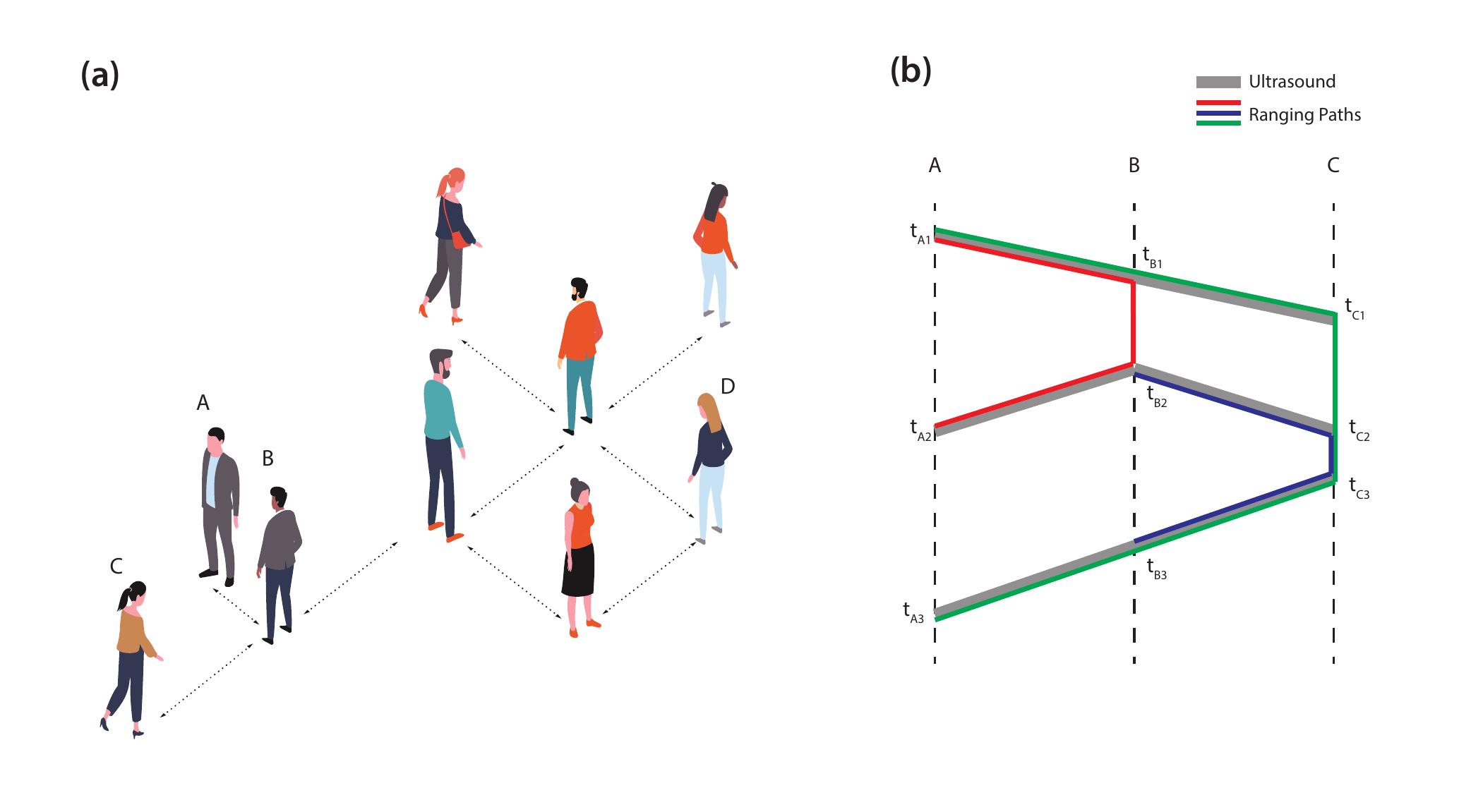}
    \caption{(a) A representative network of users. Users A, B, C are close. Users B and D are connected but not close. Users A and B are likely breaching the tolerance distance. (b) Parallel ranging. The three vertical dashed lines indicate time running down in the device's specific clock. The thick gray lines indicate a ultrasound pulse passing between devices. The red, green and blue lines highlight the ultrasound path that is used to estimate the pairwise time of flights.}
    \label{fig:ranging}
\end{figure}

\emph{Ranging}

Our ranging algorithm is inspired by Symmetrical Double Sided Two-way Ranging. The classical implementation does not require a centralised clock, and can account for processing latency if the device can accurately determine the relative time between when it heard and played the audio pulses.

For simplicity, consider a group of 3 close devices (\Cref{fig:ranging}b). Data and messages are communicated over Bluetooth, and the range is determined by Time of Flight (TOF) of ultrasound pulses. Device A sends a preparation message to all devices it is connected to, and then sends an audio signal which is heard by devices B and C. Timestamps are exchanged before the next device (B or C) takes a turn to generate their audio signal. After each pulse, each device checks to see if it has the necessary timestamps to determine a distance. For instance, the distance between A and C $d_{AC}$ is
\begin{equation*}
    d_{AC} = c \frac{(t_{C1} - t_{A1}) + (t_{A3}-t_{C3})}{2},
\end{equation*}
where $c=343$~m/s is the speed of sound. While correcting the speed of sound for local temperature and humidity could increase the accuracy \cite{Peng2007}, this information is not readily available on most regular smartphones.

As each device only needs to send the ultrasound pulse once, the total time to update scales with $T\sim|C|t/2$.\footnote{The half indicates that the $t$ represents time for two ultrasound pulses, where in our algorithm only one ultrasound pulse is needed per step. For our implementation, $t/2=0.2$~seconds.} 

To be able to accurately estimate the TOF, we must have that the devices not move during a pairwise ranging. For instance, Devices A and C (\cref{fig:ranging}) must not have moved between $t_{A1}$ and $t_{A3}$, which can be as large as $T$. In practice, if the update frequency is high enough the distance measurements are reasonably accurate.

\emph{Multiple Access Scheme}

In order to ensure that multiple devices can effectively communicate simultaneously using the Audio space, an appropriate multiple access scheme must be developed. The scheme must operate in parallel across different physical regions but has access to a very small frequency space. Carrier Sense Multiple Access (CSMA) schemes like CAN bus, over audio would be inefficient due to the range and variance of audio processing latency (8-200~ms) and low data rates.

As such, we implemented a modified hybrid Time Division (TDMA) and Space Division (SDMA) Multiple Access scheme, coordinated over Bluetooth Low Energy between neighbors. Allocations are determined live, and are specific to small physical region. This allows the ranging to operate in parallel across many clusters of devices in the network, despite using the same channel in the same frequency space.

Together, we have a simple protocol for decentralized, parallel ranging across multiple devices. It allows for users to dynamically join and leave, and scales well with user population and population density. In the next section, we describe the audio pulse and the robust detection algorithm.

\subsection{Audio Signal Processing}

\emph{Pulse Design}

As scaling considerations show, reducing $t$, the time for a single ranging operation is important to increasing the update frequency. Since sound takes 6~ms to travel 2~meters, the ranging time is dominated by the pulse duration. We do not require uniquely identifiable audio pulses (coordination happens over Bluetooth), but are limited by the performance of modern smartphones. Most smartphones operate at 44.1~or~48~kHz, and therefore our Nyquist frequency is limited to 22~kHz. However, most microphones we tested suffer a sharp decline in sensitivity above 20~kHz. 

We use simple, single bit pulses. Each audio pulse refers to a pair of 10~millisecond pulses at 18.5 and 19.25~kHz, separated by 10~ms (441 samples) and modulated by the Approximate Confined Gaussian Window \cite{Starosielec2014}. This was empirically deemed a suitable balance between pulse energy, duration and clarity using the microphones and speakers of regular smartphones. While this is lower than human hearing range, upto 20~kHz \cite{rosen2011signals}, these very short pulses were not audible to any of the researchers or users.

\emph{Pulse Detection}

Beyond good positive detection rate and false positive rejection rates, we require our pulse detection system to: (1)~accurately extract timestamps ($\pm 0.3$~ms for $\pm 10$~cm precision), (2)~reject reflected and multi-path pulses, (3)~be insensitive to pulse amplitude (to be able to recognise a self-pulse and a pulse from a device behind an obstacle), (4)~be real-time with minimal power consumption.

Since the pulse to detect is known a priori, a matched filter would be the obvious choice. However, it is unable to deal with multi-path effects: to accurately estimate the direct distance between two devices, we must ignore sound that echos off walls, the floor or other objects in an environment.

To reject reflected signals, we only consider the earliest detection. This means we will always measure the distance of the shortest successful audio path between the devices. In cases of partial obstructions, we find that these paths are changing dynamically between measurements due to micro adjustments in the environment. This has the effect of increasing the time-to-alert of the device (as the shortest paths are not measured 100\% of the time), but was found to cause minimal negative impact on the distance accuracy of alerts (\cref{tbl:accuracy}). 

Our recogniser is capable of detecting signals of amplitudes between at least $1$ and $10^4$ arbitrary units, and therefore capable of detecting pulses from a device 2~meters away through a pair of jeans, and from the device's own speakers.

Naturally, our measurement method fails for total obstructions, for instance if two devices are in two separate rooms or obstructed by a large solid object. This is in fact beneficial, since COVID-19 cannot be transmitted through such obstructions. 

Our pulses are 30~ms long, and the sound travel times are less than 20~ms, which suggests that $t/2=0.05$~seconds is the shortest ranging time possible. However, accounting for the communication overhead and reduce power consumption, we can maintain a ranging once every 0.2~seconds. 

Next, the performance of our app is discussed.

\section{Performance}
\label{sec:performance}
\FloatBarrier
To demonstrate the efficacy of the algorithm presented above, we perform three tests. First, we determine the distance estimation error for stationary pair of devices, unobstructed, partially obstructed and through clothes. Second, we determine the distance at which a user is alerted as they walk towards a device. These tests were performed by an independent testing agency by multiple different testers, and different smartphone combinations. Third, we demonstrate that the update frequency remains high in a variety of situations including with multiple devices close to each other.

\subsection{Distance Measurement Accuracy}

\begin{figure}[tb]
    \centering
    \includegraphics[width=\linewidth]{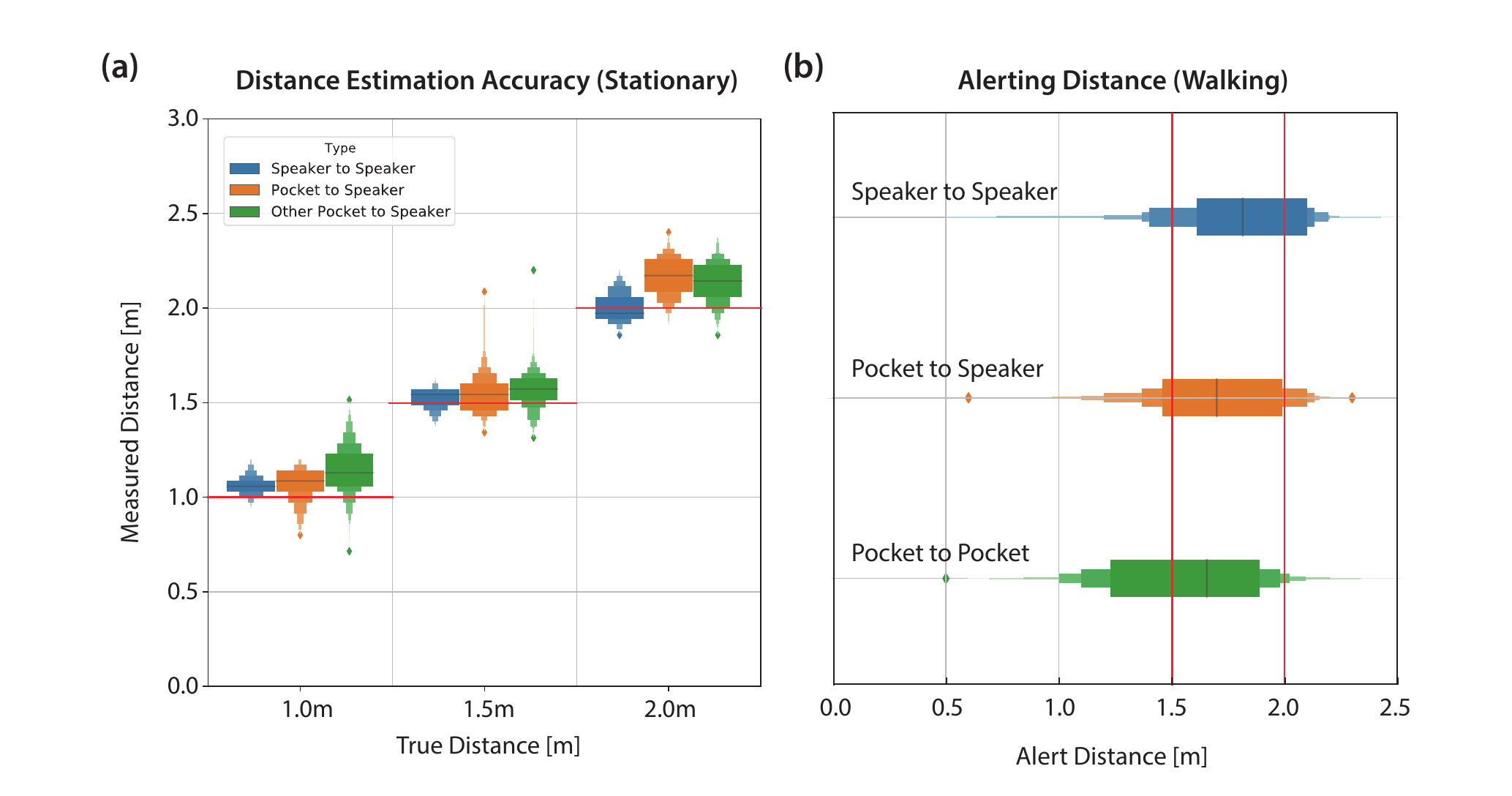}
    \caption{\textbf{Performance Summary} (a) Distribution of estimated distances for stationary phones (b) Furthest distance that the alert is sounded, as a device approaches the other starting 3~meters apart. The alert is set to sound between 1.5~and~2~meters.}
    \label{fig:perfSummary}
\end{figure}
\begin{table}[tb]
\centering
\caption{Median Average Deviation of estimated distance from true distance [meters] }
\begin{tabular}{@{}c|rrr|r@{}}
\toprule
\multirow{2}{*}{Type} & \multicolumn{3}{c|}{True Distance [m]} & \multicolumn{1}{r}{\multirow{2}{*}{\textbf{All}}} \\
 & 1~m & 1.5~m & 2~m & \multicolumn{1}{l}{} \\ \midrule
Speaker to Speaker & 0.0576 & 0.0435 & 0.0564 & \textbf{0.0564} \\
Pocket to Speaker & 0.1000 & 0.0709 & 0.1723 & \textbf{0.1006} \\
Other Pocket to Speaker & 0.1433 & 0.1006 & 0.1437 & \textbf{0.1151} \\ \midrule
\textbf{All} & \textbf{0.0861} & \textbf{0.0709} & \textbf{0.01151} & \textbf{0.0849} \\ \bottomrule
\end{tabular}
\label{tbl:accuracy}
\end{table}

The ranging accuracy was determined by holding two devices 1, 1.5 and 2~meters apart and logging the measured distance. This test was performed across two different pairs of iPhones ( [iPhone 11, iPhone SE 2016] and [iPhone 6s, iPhone 8]) and in three different configurations: (1) \emph{Speaker to Speaker}: unobstructed and with the speakers and mics facing each other, (2) \emph{Pocket to Speaker}: with one of the devices in the front pocket of a pair of jeans and (3) \emph{Other Pocket to Speaker}: with one of the phones in the back pocket.

The measurement accuracy has a median absolute deviation (MAD) of 8.5~cm, defined as the median absolute error between the measured distance and the true distance, not between the measured distance and the mean measured distance (\cref{tbl:accuracy}). The speaker to speaker measurements are more accurate, with a MAD of 5.6~cm, while introducing obstructions reduces the accuracy to 11.5~cm. This is to be expected, as the probability of only detecting a reflected path increases with obstructions, and the sound power level reduces, making a pulse harder to distinguish from noise. While multi-path has increased the variance of measurement accuracy, this represents, to the best of the authors knowledge, the best performance reported in literature for similar ranging methods. The distribution of measurements is visualized in~\cref{fig:perfSummary}a.

\subsection{Realistic Alerting Performance in Unstructured Tests}

Next, we must characterize the responsiveness of the measurement system. A test to emulate a real-world scenario is used. A tester, holding a phone, starts over 3~meters away from a stationary phone. They start walking towards the phone, until the phone alerts that the users are too close.  We record the furthermost distance at which this happens as the \emph{Alert Distance}. The breaching distance is defined as 2~meters for these tests. These tests were performed by an external testing agency, by 13~individuals with 7 different phones.\footnote{Phones used for testing: iPhone SE (2016), iPhone 6, iPhone 6S, iPhone 7, iPhone 8, iPhone 11 Pro, iPhone XR. Compatibility across more devices have been tested, including Android devices. The app requires iOS 11 and up. V1.3.2 of the application was used for this testing. All other tests were performed on V1.4}

As before, the tests were also performed in three configurations: (1) \emph{Speaker to Speaker}, as per (1) above, (2) \emph{Pocket to Speaker}, as per (2) above and (3) \emph{Pocket to Pocket} where both phones are in pockets, emulating the case of two individuals walking past each other in a hallway.  A range of materials were tested, including jeans, cotton, chiffon, silk, leather, and some tests used backpacks instead of pockets. 

\Cref{fig:perfSummary}b shows the distribution of the alerting distance for each of these different test types.\footnote{Results from Tester 13 were removed as he did not comply with test protocols. 4 of the 352 measurements (1.6\%) failed to alert before 0.5~meters.}  The mean and median alerting distancing are 1.70~and~1.76~meters respectively, with a standard deviation of 0.37~meters. The alerts occurred later for the pocket to pocket testing (median alerting distance of 1.60~meters), for the reasons mentioned earlier.
    
The application was also tested for robustness in a number of of `noisy' environments, including with music playing, in the London Underground, near construction sites and next to a highway. None hindered our application from working accurately. While noisy to our ears, these do not seem to produce enough noise in the ultrasound to confuse our audio stack. The performance (alerting rate) is marginally poorer in windy environments. The specificity of the pulse (in both the time and frequency domains) helps reject most ultrasonic noise.

\subsection{Time To Alert Tests}
\label{sec:timeToAlert}

\begin{table}[tb]
\centering
\caption{Time to Alert in seconds. Refer to \cref{sec:timeToAlert} for test description.}
\begin{tabular}{@{}rrrrr@{}}
\toprule
Test Number & (1) & (2) & (3) & (4) \\
Type & 2 Devices & 2 Devices & 4 Devices (edge test) & 4 Devices (center test)\\
Mode & Speaker to Speaker & Pocket to Speaker & Pocket to Speaker & Pocket to Speaker\\ \midrule
Median & 0.45 & 0.92 & 2.07 & 2.83 \\
Mean & 0.48 & 0.85 & 2.24 & 3.06 \\
Avg. Deviation & 0.07 & 0.22 & 0.96 & 0.79 \\
Min & 0.37 & 0.30 & 0.17 & 1.63 \\
Max & 0.63 & 1.20 & 4.12 & 5.12 \\ \bottomrule
\end{tabular}
\label{tbl:timeToAlert}
\end{table}

To demonstrate the app's responsiveness, we performed a number of tests with multiple phones, initially at 2.25~m apart, outdoors. One phone is brought to within 1.75~m of the other(s) such that it breaches the 2~m threshold. The time between crossing the 2~m threshold and receiving an alert is measured. This test concept is repeated in 4 situations. (1)~Two phones (SE,~5S) where they are oriented speaker to speaker, (2) Two phones (SE,~5S) where one is in a pocket, (3) Four phones (SE,~5S,~11,~XSMax) arranged in a square, where one device is moved along one edge (alerting a single other device), and (4) Four phones (SE,~5S,~11,~XSMax) arranged in a square where one device is moved into the center (alerting all the other devices). The stationary devices were placed on a surface, and the last was in the tester's denim short pockets. Links to videos of the tests can be found here.\footnote{  Two device tests \url{https://youtu.be/9wfNRw0VPDg}. Four device tests \url{https://youtu.be/4xXfuktKIaE}. }

We observe a mean time to alert of 0.48~seconds in the unobstructed case with two devices, which corresponds to two audio pulses being transmitted (\Cref{tbl:timeToAlert}, test 1). When the app is used through pockets, the responsiveness reduces to a 0.85~seconds (test 2). This is due to missed readings caused by the the obstruction. The minimum time to alert is still fast, at just 0.3~seconds. We note the fraction of missed detection is dependent on the relative positioning and environmental factors. With four devices, the time to alert is roughly doubled, not quadrupled (test 3), due to our parallel ranging architecture.

These alerting times were considered sufficient for the purpose of social distancing - individuals walking past each other do not need to be alerted. If the application requires it, responsiveness can be increased, albeit with greater battery consumption.

\section{Conclusion}
\label{sec:conclusion}

In this implementation, we were able to achieve median estimation absolute errors of 8.5~cm without averaging multiple readings and across different device types. The application provides effective real-time alerts in all the tested situations, including when obstructed by clothes and users. In best and worst case environments the alerting time was found to be on average 0.48-seconds and 3.06-seconds respectively. Our application works using standard smartphone technologies. No beacons, calibration or mapping are required, and the system scales well with device count and density. The application provides complete user privacy, and there is no centralized communication or coordination. 

Beyond the technical merits of our work, we believe our application serves an important role to make it easier and more convenient for individuals to maintain social distancing guidelines. 

\section*{User Data Privacy Note}
The entire algorithm provides complete user anonymity. The algorithm only processes sounds above 18~kHz (well above any frequency associated with speech), which are never stored for more than 1~second. The app requires no internet connection (apart from a login process which authenticates the user to use the app). The app does not require GPS or any location services. The app does not store or expose any personal information about any of its users. These data privacy notes were independently verified by researchers at the University of Birmingham.

\section*{Data and Code Availability}

The empirical data presented in this paper and the code to produce the graph is available at Github (\url{https://github.com/dev10110/MindTheGap-Paper}). Please contact River Baig (\url{mailto:river@hackpartners.com}) for the source code. Visit \url{mindthegap.today} to obtain the app. 

\section*{Acknowledgements}

The authors would like the thank Hack Partners Limited for financially supporting this project. In addition, we would like to thank Colin Dente for insightful discussions about the design of the audio processing stack.

\bibliographystyle{unsrt}  


\begin{thebibliography}{10}

\bibitem{WHO2020}
\newblock {Overview of public health and social measures in the context of
  COVID-19}.
\newblock {\em World Health Organization 2020.}, May 2020.

\bibitem{UKGuidance}
GOV.UK.
\newblock {Staying Alert and Safe Social Distancing after 4 July}, 2020.
\newblock \url{https://www.gov.uk/government/publications/staying-alert-and-safe-social-distancing/staying-alert-and-safe-social-distancing-after-4-july}.

\bibitem{MendozaSilva2019}
Germ{\'{a}}n~Mart{\'{\i}}n Mendoza-Silva, Joaqu{\'{\i}}n Torres-Sospedra, and
  Joaqu{\'{\i}}n Huerta.
\newblock A meta-review of indoor positioning systems.
\newblock {\em Sensors}, 19(20):4507, October 2019.

\bibitem{Ward1997}
Andy Ward, Alan Jones, and Andy Hopper.
\newblock {A new location technique for the active office}.
\newblock {\em IEEE Personal Communications}, 4(5):42--47, 1997.

\bibitem{Cricket}
Nissanka~B. Priyantha, Anit Chakraborty, and Hari Balakrishnan.
\newblock The cricket location-support system.
\newblock In {\em Proceedings of the 6th Annual International Conference on
  Mobile Computing and Networking}, MobiCom '00, page 32–43, New York, NY,
  USA, 2000. Association for Computing Machinery.

\bibitem{Fukuju}
Y.~Fukuju, M.~Minami, H.~Morikawa, and T.~Aoyama.
\newblock {DOLPHIN}: an autonomous indoor positioning system in ubiquitous
  computing environment.
\newblock In {\em Proceedings {IEEE} Workshop on Software Technologies for
  Future Embedded Systems. {WSTFES} 2003}. {IEEE} Comput. Soc, 2003.

\bibitem{Peng2007}
Chunyi Peng, Guobin Shen, Yongguang Zhang, Yanlin Li, and Kun Tan.
\newblock {BeepBeep}.
\newblock In {\em Proceedings of the 5th international conference on Embedded
  networked sensor systems - {SenSys07}}. {ACM} Press, 2007.

\bibitem{Urena2018}
Jes{\'{u}}s Ure{\~{n}}a, {\'{A}}lvaro Hern{\'{a}}ndez, J.~Jes{\'{u}}s
  Garc{\'{i}}a, Jos{\'{e}}~M. Villadangos, M.~Carmen P{\'{e}}rez, David Gualda,
  Fernando~J. {\'{A}}lvarez, and Teodoro Aguilera.
\newblock {Acoustic Local Positioning with Encoded Emission Beacons}.
\newblock {\em Proceedings of the IEEE}, 106(6):1042--1062, 2018.

\bibitem{Starosielec2014}
Sebastian Starosielec and Daniel H{\"{a}}gele.
\newblock {Discrete-time windows with minimal RMS bandwidth for given RMS
  temporal width}.
\newblock {\em Signal Processing}, 102:240--246, 2014.

\bibitem{rosen2011signals}
Stuart Rosen and Peter Howell.
\newblock {\em Signals and systems for speech and hearing}, volume~29.
\newblock Brill, 2011.

\end{thebibliography}

\end{document}